\documentclass[12pt]{iopart}

\usepackage{iopams}
\usepackage{graphicx}
\usepackage{cite}
\usepackage{hyperref}
\usepackage{xcolor}
\usepackage{caption}

\newcommand{\ket}[1]{\left|#1\right\rangle}
\newcommand{\bra}[1]{\left\langle#1\right|}
\newcommand{\mel}[3]{\left\langle#1\right|#2\left|#3\right\rangle}

\begin{document}

\title{Boundary-localized non-Hermitian modes in an absorbing quantum walk}

\author{Francisco Riberi}

\address{
Department of Electrical and Computer Engineering,
University of New Mexico,
NM 87131, USA
}

\address{Center for Quantum Information and Control, University of New Mexico, NM 87131, USA}

\ead{friberi@unm.edu}
\renewenvironment{abstract}
{\par\noindent\textbf{Abstract.}\ }
{\par\vspace{1em}}
\begin{abstract}
    We introduce and solve from first principles a continuous-time quantum walk with absorption generated by a Lindblad boundary sink of arbitrary strength. Projecting onto the surviving walker sector maps the problem to a
non-Hermitian tight-binding Hamiltonian with a rank-one imaginary defect
on the semi-infinite line. We obtain closed-form expressions for the exact propagator, from which the first-passage statistics follow.  Weak coupling limits absorption  through inefficient transfer into the sink, whereas for strong dissipation, occupation at the boundary is suppressed by Zeno-like reflection, accompanied by the emergence of a localized non-Hermitian mode. Despite the different physical origins of these suppression mechanisms,
their respective absorption probabilities exhibit an asymptotic duality. The evolution is conveniently visualized in phase space, where the localized mode has an exponentially confined Wigner representation near the edge site.
    \end{abstract}
    \maketitle

    \section{Introduction}

    Quantum walks (QWs) \cite{aharonov1993,kempe2009,Venegas2012,portugal2013,kadian2021}
    provide a canonical framework to study coherent transport on discrete systems and play a central role in quantum dynamics, simulation, and algorithms \cite{ambainis2003,shenvi2003,aharonov2009,childs2009}.  Both coined \cite{aharonov2001,ambainis2001} and continuous-time \cite{farhi1998,childs2004,mulken2011} variants exhibit genuine quantum features such as ballistic spreading and interference, leading to behavior with no classical analogue \cite{Childs2003}. These properties have motivated extensive theoretical \cite{brun2003b,brun2003,konno2005,strauch2006,yin2008,whitfield2010,caceres2010,nizama2012,attal2012open,nizama2019,chen2025,king2026} and experimental \cite{travaglione2002,du2003,zahringer2010,qiang2021,razzoli2024} investigations.
    
    A central question in classical \cite{siegert1951} and quantum \cite{friedman2017,ruiz2023} transport is that of first-passage or hitting times, which characterize arrival statistics at a target set of sites. In QWs, absorption and detection are inherently nonunitary, and different operational definitions
    can produce qualitatively distinct behaviors
\cite{varbanov2008,magniez2012,ruiz2023}. Even in one-dimensional geometries, absorbing QWs exhibit phenomena absent in classical first-passage processes, including partial reflection, incomplete absorption, and strong sensitivity to the initial state \cite{bach2004,konno2003,kuklinski2018,kuklinski2018b}.
    
 Many existing treatments of absorbing QWs implement absorption
through repeated projective measurements performed at
discrete times \cite{ambainis2001,yamasaki2003,bach2004,kwek2011}. The walk is terminated upon successful detection. In these models, the first-passage statistics depend not only on the walk itself, but also on the measurement scheme: the sampling
interval constitutes an additional parameter, whereas null measurements project and
renormalize the surviving state.
Continuous-time stochastic-detection protocols avoid a fixed sampling
interval but likewise introduce an external monitoring process whose
statistics enter the first-detection problem
\cite{varbanov2008,ruiz2023}. Consequently, phenomena such as Zeno
suppression depend explicitly on how and when
the walker is interrogated.

Here instead, we model absorption as an autonomous loss process,
generated by a Lindblad sink coupled locally to the boundary. No external detection is assumed. First-passage statistics follow
directly from the probability current into the sink, with a single physical 
parameter describing the competition between coherent
hopping and irreversible loss. Restricting the dynamics to the surviving walker sector yields an exact non-Hermitian Hamiltonian with a localized imaginary defect. This non-Hermitian
description is not imposed phenomenologically, but emerges from the
trace-preserving evolution of the enlarged system. The model is relevant to engineered quantum transport platforms with controllable dissipation, including photonic waveguide arrays \cite{biggerstaff2015}, superconducting circuits \cite{sheremet2023}, trapped ions \cite{bruzewicz2019}, and cold-atom quantum simulators \cite{damanet2019}.
    
The evolution of the walker on the semi-infinite line admits an exact
solution, yielding a closed-form propagator from which the survival
probability and first-passage density follow exactly. The dynamics show a competition between coherent transport and boundary loss. When the absorption rate exceeds the coherent hopping rate, access to the sink becomes increasingly suppressed due to dissipative reflection, accompanied by the emergence of a localized non-Hermitian mode \cite{borgnia2020non}. The crossover admits a direct phase-space interpretation: the boundary mode has an exponentially localized Wigner function, which we term a Wigner droplet. Remarkably, as the initial state is moved away from the boundary, the absorption probability becomes asymptotically invariant under the exchange \((\kappa/\Omega)\leftrightarrow(\Omega/\kappa)\). This shows that weak coupling and strong dissipation can produce identical long-time absorption through distinct physical mechanisms. 
\begin{figure*}[t!]
    \centering
    \includegraphics[width=\textwidth]{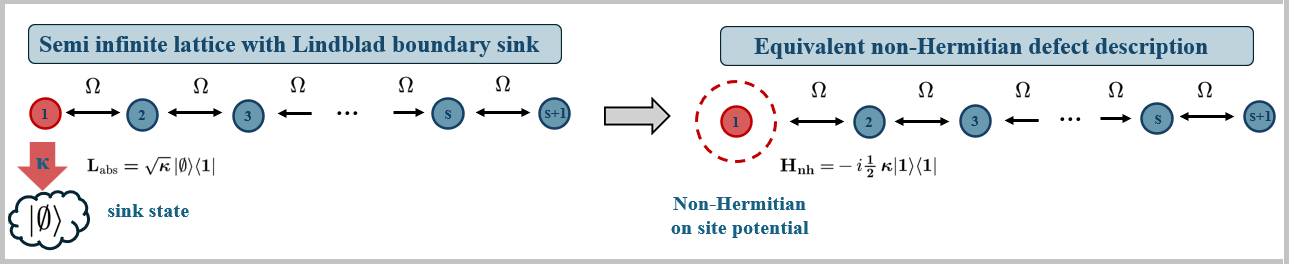}
    \vspace*{-5mm}
    \captionsetup{width=\textwidth}
    \caption{
    \textbf{Schematic of the boundary-absorption model and its effective non-Hermitian description.}
    A particle propagates coherently on a semi-infinite lattice with hopping rate $\Omega$, while a Lindblad sink irreversibly removes population from the boundary site $s=1$ at rate $\kappa$ (left).
    This loss is equivalently described by the non-Hermitian boundary defect
    $-i(\kappa/2)\ket{1}\!\bra{1}$ in the effective Hamiltonian (right).
    }
    \label{diagram}
\end{figure*}
    
 \section{Model} 
     
    We consider a continuous-time QW  on the semi-infinite lattice \(s\in\mathbb N\), with coherent hopping and absorption localized at the boundary. Evolution is generated by a nearest-neighbor tight-binding Hamiltonian with hopping rate $\Omega$, 
    \begin{eqnarray} 
    H_+ 
    &=& 
    \Omega \sum_{s\ge1} \ket{s}\bra{s} 
    - 
    \frac{\Omega}{2} 
    \sum_{s\ge1} 
    \left( 
    \ket{s}\bra{s+1} 
    + 
    \ket{s+1}\bra{s} 
    \right), 
    \end{eqnarray} 
    which produces ballistic propagation on the half-line. 
    Absorption is included by coupling the edge site $s=1$ to a sink state $\ket{\emptyset}$ through the jump operator 
    $L_{\rm abs} 
    = 
    \sqrt{\kappa}\,\ket{\emptyset}\bra{1},$ 
    where $\kappa$ is the absorption strength. The full density matrix obeys the following Lindblad master equation: 
    \begin{eqnarray} 
    \dot\rho 
    &=& 
    -i[H_+,\rho] 
    + 
    L_{\rm abs}\rho L_{\rm abs}^\dagger 
    - 
    \frac{1}{2}\{L_{\rm abs}^\dagger L_{\rm abs},\rho\}. 
    \end{eqnarray} 
    Projecting onto the surviving walker sector yields a trace-decreasing 
evolution governed by the effective non-Hermitian Hamiltonian $H_{\rm eff} 
    = 
    H_+ 
    - 
    \frac{i\kappa}{2}\ket{1}\bra{1}$, as illustrated in Fig. \ref{diagram}: 
    \begin{eqnarray} 
     \dot \rho_+&=& -iH_{\rm eff}\rho_+ 
    +i\rho_+H_{\rm eff}^\dagger, \qquad \rho_+=\Pi_+ \rho\Pi_+ , \label{ev} 
    \end{eqnarray} 
    with $\Pi_+=I-|\emptyset\rangle \langle\emptyset|$. 
    Thus the absorbing QW reduces  to a tight-binding model with a rank-one imaginary defect. 
    The defect continuously removes probability amplitude reaching the boundary site. 
    Evolution is determined by the  propagator 
    $K_\kappa(s,s';t) 
    = 
    \mel{s}{e^{-iH_{\rm eff}t}}{s'},$ 
    and the reduced density matrix evolves as 
    \begin{eqnarray} 
    \rho_{ss'}^{(+)}(t) 
    &=& 
    \sum_{u,v\ge1} 
    K_\kappa(s,u;t)\, 
    \rho_{uv}^{(+)}(0)\, 
    K_\kappa^*(s',v;t). \label{reduced} 
    \end{eqnarray}

    \section{Analytic solution} 
     
    We solve exactly for the propagator of the non-Hermitian Hamiltonian 
\(H_{\rm eff}\), from which the relevant dynamical observables can be 
obtained. In the absence of absorption (\(\kappa=0\)), the system 
reduces to a coherent QW on the half-line with edge site \(s=1\) and a 
hard-wall boundary condition at \(s=0\). The corresponding propagator 
follows from the method of images \cite{chandrasekhar1943} applied to 
the infinite-lattice solution \cite{nizama2012}, 
    \begin{eqnarray} 
    K_D(s,s_0;t) 
    &=& 
    e^{-i\Omega t} 
    \left[ 
    i^{s-s_0}J_{s-s_0}(x) 
    - 
    i^{s+s_0}J_{s+s_0}(x) 
    \right], \label{KD} 
    \end{eqnarray} 
    with $x=\Omega t$. The second term corresponds to the image contribution and enforces the hard-wall boundary condition \(K_D(0,s_0;t)=0\). Absorption can be incorporated exactly through the Green functions of 
the effective and reflecting Hamiltonians, 
\begin{eqnarray} 
G_\kappa(z)&=&\frac{1}{z-H_{\rm eff}}, 
\qquad 
G_D(z)=\frac{1}{z-H_+}. 
\end{eqnarray} 
Since the absorber acts only on the boundary site, it enters as the 
rank-one non-Hermitian defect 
\( 
V_{\rm abs}=-\frac{i\kappa}{2}\ket{1}\bra{1}. 
\) 
Thus, its repeated scattering contributions can be resummed exactly 
(see \ref{app:laplace}). 
Propagation from \(s_0\) to \(s\) decomposes into a direct bulk term and processes involving repeated scattering from the absorbing boundary, 
    \begin{eqnarray} 
    G_\kappa(s,s_0;z) 
    &=& 
    G_D(s,s_0;z) 
    - 
    \frac{i\kappa}{2} 
    \frac{ 
    G_D(s,1;z)\,G_D(1,s_0;z) 
    }{ 
    1+\frac{i\kappa}{2}G_D(1,1;z) 
    }. 
    \end{eqnarray} 
    We compute the bulk contribution in terms of the solution for the unconstrained walk on a line,  $G_D(s,s_0;z)= g_{s-s_0}(z)-g_{s+s_0}(z),$ with  
    \begin{eqnarray} 
        g_n(z) 
    &=& 
    \int_{-\pi}^{\pi} 
    \frac{dk}{2\pi} 
    \frac{e^{ikn}} 
    {z-\Omega(1-\cos k)}= 
    -\frac{2}{\Omega} 
    \frac{q^{|n|+1}}{1-q^2}. \label{gn} 
    \end{eqnarray} 
    Here $n$ denotes the separation on the infinite lattice, and $q$ is defined implicitly through $z=\Omega(1-\frac{1}{2}(q+q^{-1}))$. 
    Substituting these expressions into $G_{\kappa}(s,s_0; z)$ gives the exact Green function: 
    \begin{eqnarray} 
    G_\kappa(s,s_0;z) 
    &=& 
    G_D(s,s_0;z) 
    - 
    \frac{2i\eta}{\Omega} 
    \frac{q^{s+s_0}}{1-i\eta q}, \label{Gk} 
    \end{eqnarray} 
    with $\eta=\kappa/\Omega$ a \emph{dimensionless absorption strength} that quantifies the competition between dissipation and coherent hopping.  
    We obtain the propagator by inverse Laplace transform, 
    \begin{eqnarray} 
    K_\kappa(s,s_0;t) 
    &=& 
    \frac{1}{2\pi i} 
    \int_{\mathcal C} 
    dz\, 
    e^{-izt} 
    G_\kappa(s,s_0;z), \label{invLap} 
    \end{eqnarray} 
    where $\mathcal{C}$ is a contour running horizontally in the complex $z$-plane above all poles and branch cuts of $G_\kappa(s,s_0;z)$. 
    The inverse transform of $G_D(s,s_0;z)$ reproduces the hard-wall propagator in Eq.~(\ref{KD}), while the scattering contribution encodes the absorber-induced boundary effects.  
     
    For \(\eta<1\), the denominator in the second term of Eq.~(\ref{Gk}) admits the convergent geometric expansion 
    \( 
    (1-i\eta q)^{-1} 
    = 
    \sum_{r=0}^\infty (i\eta q)^r. 
    \) 
    Computing the propagator reduces to a series of inverse transforms of powers \(q^{\ell+r}\), which can be obtained using the known relationship \(G_D(r,1;z)\propto q^r\). The result is: 
    \begin{eqnarray} 
    K^{<}_\kappa(s,s_0;t) 
    &=& 
    K_D(s,s_0;t) 
    + 
    2\eta\, i^{\ell} e^{-i\Omega t} 
    \sum_{r=0}^{\infty} 
    (-\eta)^r 
    \frac{\ell+r}{x} 
    J_{\ell+r}(x), 
    \label{Bessel} 
    \end{eqnarray} 
    where $\ell=s+s_0$ and $x=\Omega t$. 
Although Eq.~(\ref{Bessel}) was derived using a geometric series that 
converges only for $\eta<1$, the resulting Bessel expansion remains valid 
at $\eta=1$---and, in fact, for any finite $\eta>0$---because, at fixed 
$x$, $J_{\nu}(x)$ decays factorially with increasing $\nu$. This provides an analytic 
continuation beyond the original derivation domain which, as shown explicitly 
in \ref{equivalence}, coincides with the exact 
finite-time propagator for all finite \(\eta>0\).

    For \(\eta>1\), the analytic structure changes qualitatively, as the denominator 
    \( 
    1-i\eta q 
    \) 
    vanishes at 
    $q_p=-i\eta^{-1},$ 
    with \(|q_p|<1\). As a result, the  contour integral used to evaluate Eq. (\ref{invLap}) encloses a pole in addition to the continuum branch-cut contribution (see \ref{app:laplace}). We derive an equivalent expression for the propagator that separates into a continuum term and a discrete boundary-localized mode: 
    \begin{eqnarray} 
    K^{>}_\kappa(s,s_0;t) 
    &=& 
    K^{\rm cont}_{\kappa}(s,s_0;t) 
    + 
    (1-q_p^2)\,q_p^{s+s_0-2}e^{-iz_p t}, \label{K2} 
    \end{eqnarray} 
    where 
    $z_p 
    = 
    \Omega 
    - 
    \frac{i\Omega}{2} 
    \left( 
    \eta-\eta^{-1} 
    \right). 
    $ 
 Note that the crossover at \(\eta=1\) is not an exceptional point: 
\(H_{\rm eff}\) exhibits no coalescence of discrete eigenvalues and 
eigenvectors, and the Green function develops no higher-order pole 
associated with a defective eigenstate. Instead, as 
\(\eta\) approaches 1 from above, the boundary-localized mode delocalizes and reaches the 
continuum threshold. For \(\eta\leq1\), it no longer defines a 
normalizable discrete eigenstate. 
 
    We evaluate the continuum contribution in Eq. (\ref{K2}) by expanding 
    \( 
    (1-i\eta e^{ik})^{-1} 
    = 
    -\sum_{r=1}^{\infty}(i\eta e^{ik})^{-r},\) 
    which converges for \(|\eta|>1\): 
    \begin{eqnarray} 
    K^{\rm cont}_\kappa(s,s_0;t) 
    &=& 
    K_D(s,s_0;t) 
    + 
    2 i^{\ell} e^{-i\Omega t} 
    \sum_{r=1}^{\infty} 
    (-1)^{r+1} \eta^{1-r} 
    \frac{\ell-r}{x} 
    J_{\ell-r}(x). 
    \label{Bessel2} 
    \end{eqnarray} 
As for Eq.~(\ref{Bessel}), the inverse-power expansion used to derive 
Eq.~(\ref{Bessel2}) converges only in its original domain, here 
$\eta>1$. Nevertheless, the resulting Bessel series and the accompanying pole term 
admit analytic continuation to any finite \(\eta>0\). \ref{equivalence} 
proves explicitly that their sum reproduces the same exact finite-time 
propagator as Eq.~(\ref{Bessel}). 
For $\eta<1$, however, Eq. (\ref{K2}) represents only a rearrangement of the terms in the infinite series conforming the propagator. The continued ``pole'' term is only formal: 
since $|q_p|=\eta^{-1}>1$, it does not represent a normalizable 
boundary-localized eigenmode.
   \section{Discussion}

      \begin{figure*}[t!]
    \centering
    \includegraphics[width=\textwidth]{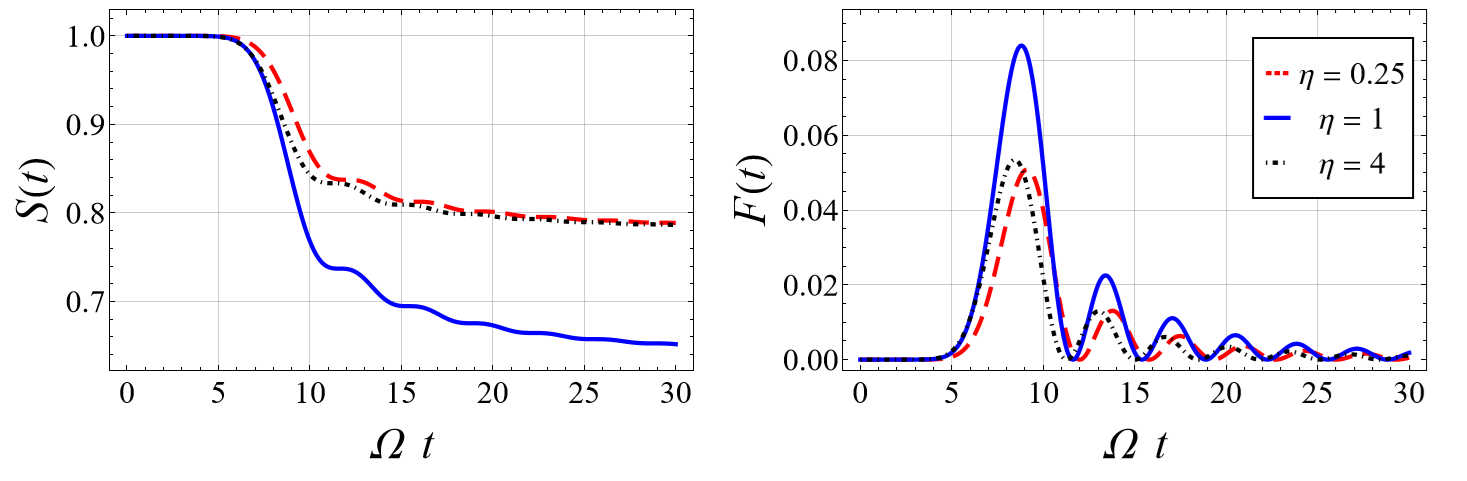}
    \vspace*{-7mm}
    \caption{
\textbf{Survival probability and first-passage density for an absorbing quantum walk and localized initial condition.}
The curves show weak absorption (\(\eta=0.25\), red, dashed), at the crossover (\(\eta=1.0\), solid, blue), and strong dissipation (\(\eta=4\), black dot-dashed).
Left: Survival probability \(S(t)\), defined as the probability that the walker remains in the semi-infinite lattice up to time \(t\). Note the  nearly identical \(S(t)\) asymptotics for the \(\eta=0.25\) and \(\eta=4\) cases. This shows that the duality-breaking finite-distance corrections in Eq. (\ref{PabsExact}) are
already negligible for our choice of $s_0$.}
Right: First-passage density \(F(t)=\kappa |\psi_1(t)|^2\), giving the probability density for absorption at the boundary. 
  The oscillations in \(F(t)\) reflect coherent interference between amplitudes arriving at the boundary through direct and boundary-reflected propagation.
Near \(\eta=1\), these competing effects are optimally balanced and
absorption is most efficient; the optimum approaches \(\eta=1\) in the
distant-initial-state limit.
Parameters: \(s_0=8\), \(\Omega=1\), and \(0\le\Omega t\le30\).
    \label{SF}
    \end{figure*}
    
    The exact propagator determines all dynamical observables of the absorbing QW. For an initial state localized at \(s_0\), the survival probability is
    \begin{eqnarray}
    S(t|s_0)
    &=&
    \sum_{s\ge1}
    |K_\kappa(s,s_0;t)|^2, \label{S}
    \end{eqnarray}
    while the complement $1-S(t|s_0)$ describes absorption at the boundary sink. 
    
 Using Parseval's identity to express Eq. (\ref{S}) in terms of the Green
function, we derive the following exact integral representation of the total absorption probability,
\(
P_{\rm abs}(s_0)=\lim_{t\to\infty}[1-S(t|s_0)],
\) (see \ref{app:pabs}):
    \begin{eqnarray}
   P_{\rm abs}(s_0;\eta)
&=&
\frac{2\eta}{\pi}
\int_0^\pi dk\,
\frac{\sin k}
{1+\eta^2+2\eta\sin k}
+
\frac{2\eta}{\pi}
\int_0^1 dq\,
\frac{
q^{2s_0-2}(1-q^2)
}{
1+\eta^2q^2
}.
    \label{PabsExact}
    \end{eqnarray}
The first term in Eq.~(\ref{PabsExact}) is invariant under the
\(\eta\leftrightarrow\eta^{-1}\) transformation. The second is a
finite-distance correction that vanishes as \(s_0^{-2}\). Importantly, for $s_0 \gg 1$,
its leading \(O(s_0^{-2})\) contribution is itself invariant under the exchange
\(\eta\leftrightarrow\eta^{-1}\), so the first violation of the
weak--strong symmetry appears only at \(O(s_0^{-3})\). Thus, for
\(s_0\gg1\),
\(
P_{\rm abs}(s_0;\eta)
\simeq
P_{\rm abs}(s_0;\eta^{-1}),
\)
with the symmetry becoming exact as \(s_0\to\infty\). Remarkably,
agreement is already excellent for \(s_0=8\), as shown in
Fig.~\ref{SF}. This symmetry, however, does not extend to the full
transient dynamics.
    
The origin of this weak--strong symmetry can be understood from the loss flux into the sink,
\begin{equation}
F(t|s_0)=-\dot S(t|s_0)
=\kappa\,|K_\kappa(1,s_0;t)|^2,
\qquad
P_{\rm abs}
=\int_0^\infty dt\,F(t|s_0).
\label{effective}
\end{equation}
 Equation~(\ref{effective}) tells us that two conditions are required
for effective absorption: the walker must reach the boundary with appreciable
probability $|K_\kappa(1,s_0;t)|^2$, and the sink must remove this population efficiently. For weak
absorption, the walker can readily reach the edge site, but
$\kappa\ll\Omega$ removes this population inefficiently, giving
$P_{\rm abs}\sim4\eta/\pi$ for a distant initial condition.
Conversely, in the limit of strong dissipation ($\eta\gg1$), the sink
removes amplitude so rapidly that appreciable occupation cannot build up at $s=1$. 
The boundary becomes effectively reflecting---the continuous-loss analogue
of quantum Zeno suppression---and $P_{\rm abs}\sim4/(\pi\eta)$. It is thus reasonable that these complementary
mechanisms could produce the same leading long-time absorption probability under
\(\eta\leftrightarrow\eta^{-1}\) when $s_0 \gg1$. In the distant-initial-state limit,
the competition is optimally balanced at \(\eta=1\), where the absorption
probability reaches \(1-2/\pi\).

However, for a state initialized near the absorber, the two regimes are
no longer equivalent. For \(\eta>1\), as the walker is initialized closer to the
sink, its overlap with the boundary-localized mode increases. Since the mode
is concentrated near the lossy boundary, its probability continuously
leaks into the sink. This is reflected in an imaginary component of its eigenvalue,
\(
z_p=\Omega-i\Gamma_p/2,
\)
which gives a survival probability decaying as \(e^{-\Gamma_p t}\).
Thus, initial states closer to the sink can be absorbed more efficiently
for \(\eta>1\). As \(s_0\) increases, the overlap with the localized mode is
exponentially suppressed as
\(
|\phi_p(s_0)|^2\propto\eta^{-2(s_0-1)},
\)
and the weak--strong duality is recovered.

 The transient first-passage dynamics, shown in Fig.~\ref{SF}, also distinguishes
these regimes. At short times, \(F(t|s_0)\) exhibits oscillations arising
from interference between direct and boundary-reflected propagation paths.
Near \(\eta\simeq1\), \(S(t)\) decays rapidly, as absorption is most
efficient. By contrast, for \(\eta\gg1\) and \(\eta\ll1\), the first-passage density
exhibits similar interference-induced oscillations, whose amplitude is
suppressed by Zeno reflection and inefficient absorption, respectively. This crossover admits a natural interpretation in phase
space.
\begin{figure*}[t!]
    \centering
    \includegraphics[width=\textwidth]{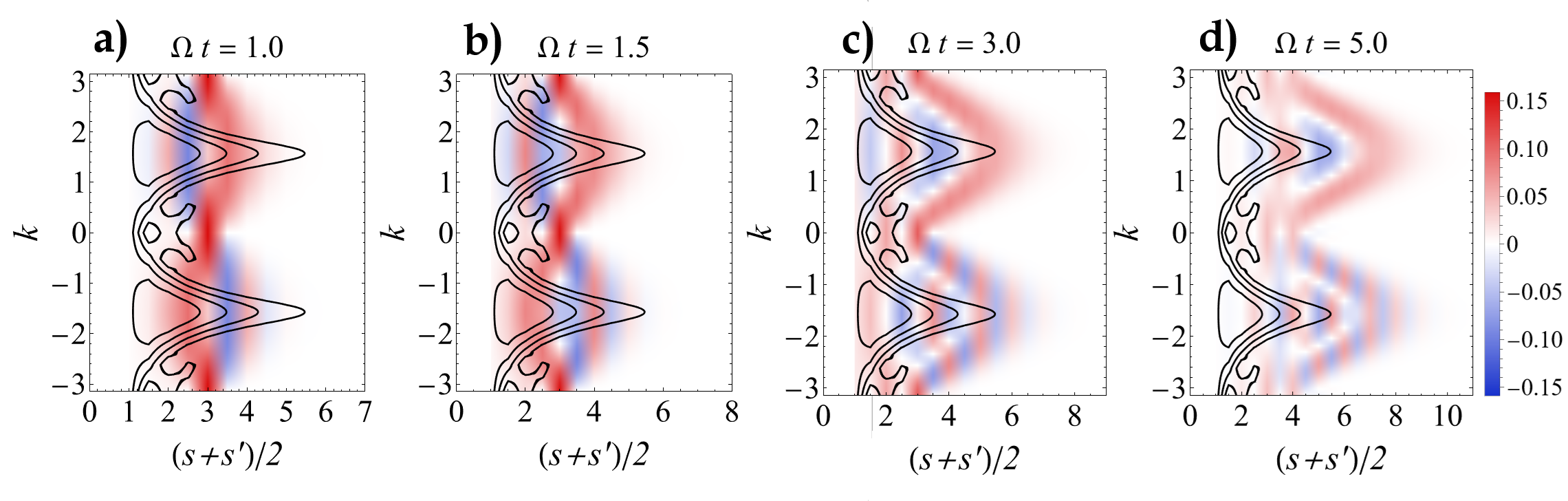}
    \vspace*{-7mm}
\caption{
    \textbf{Phase-space dynamics of a continuous-time quantum walk with an absorbing boundary.}
    Snapshots of the doubled-lattice Wigner function \(W_+(x_c,k,t)\), with
    \(x_c=(s+s')/2\), for a walker initially localized near the boundary.
    The colored density shows the physical Wigner function of the surviving walker,
    while the black contours show the Wigner droplet
    \(W_{{\rm pole},{\rm pole}}\)---the phase-space representation of the
    localized boundary mode---rescaled for visibility.
    At early times [(a)] the full wavepacket spreads ballistically and behaves
    essentially as a free quantum walk; the pole, continuum, and their interference
    contributions combine coherently to reconstruct this short-time dynamics.
    As the wavepacket reaches the boundary [(b)], loss and coherent reflection generate
    an increasingly pronounced interference structure. The phase-space fringes arise
    from interference between bulk propagation and amplitudes scattered from the
    absorbing boundary.
    At later times [(c,d)], the overall magnitude of the Wigner function decreases
    with the survival probability as the walker is absorbed, while
    \(W_{{\rm pole},{\rm pole}}\) remains spatially confined near the sink.}
    In all cases, \(\Omega=1\), \(\kappa=1.5\), and \(s_0=3\).
    \label{CV}
\end{figure*}
    
    The dynamics can be conveniently visualized in terms of the doubled-lattice Wigner function ($m \geq 2$) \cite{bizarro1994,hinarejos2012}:
    \begin{eqnarray}
    W_+(m,k,t)
    &=&
    \frac{1}{2\pi}
    \sum_{n=1}^{m-1}
    \psi_n(t)\psi_{m-n}^*(t)
    e^{-i(2n-m)k},
    \end{eqnarray}
    where \(\psi_s(t)=K_\kappa(s,s_0;t)\) for a localized initial condition. 
    Substituting the exact propagator yields an explicit Bessel-series representation of the Wigner function (see \ref{app:wigner}). For weak absorption, \(\eta<1\), the propagator separates into hard-wall and boundary-return contributions, leading to the following decomposition: 
    \begin{eqnarray}
    W_+
    &=&
    W_{DD}
    +
    (W_{DB}
    +
    W_{BD})
    +
    W_{BB}. \label{Wig}
    \end{eqnarray}
    The terms correspond, respectively, to coherent bulk propagation, bulk--boundary interference, and repeated boundary-return processes. Equation (\ref{Wig}) cleanly separates the competing effects contributing to the evolution.
    
For \( \eta>1\), it is instead natural to use the
strong-coupling propagator, Eq. (\ref{K2}),
which induces a different decomposition:
\begin{equation}
W_+
=
W_{{\rm cont},{\rm cont}}
+
\left(
W_{{\rm cont},{\rm pole}}
+
W_{{\rm pole},{\rm cont}}
\right)
+
W_{{\rm pole},{\rm pole}}.
\label{WigPole}
\end{equation}
The last term, \(W_{{\rm pole},{\rm pole}}\), is the phase-space
representation of the boundary-localized non-Hermitian mode, which we
refer to as the \emph{Wigner droplet}. The mode's wave function scales as
\(
\psi_s^{\rm pole}\propto q_p^{s-1},
\)
with \(|q_p|=\eta^{-1}\), so its probability density is
\( \propto \eta^{-2(s-1)}\). Accordingly, the droplet has an envelope
that decays exponentially with the doubled-lattice coordinate \(m\),
\begin{eqnarray}
|W_{{\rm pole},{\rm pole}}(m,k,t)|
&\sim&
e^{-\Gamma_p t}\eta^{-m},
\qquad
\Gamma_p
=
\kappa-\frac{\Omega^2}{\kappa}.
\end{eqnarray}
The boundary mode wavefunction satisfies
\(
|\psi_s^{\rm pole}|\propto\eta^{-(s-1)}
=e^{-(s-1)/\xi_{\rm loc}},
\)
with amplitude localization length
\(
\xi_{\rm loc}=1/\ln\eta.
\)  A complementary quantitative measure of this spectral localization is
the inverse participation ratio, defined as  ${\rm IPR}_{\rm pole}
=
\sum_{s=1}^{\infty}|\psi_s|^4$. For the normalized boundary mode,
\begin{equation}
{\rm IPR}_{\rm pole}
=
\sum_{s=1}^{\infty}|\psi_s^{\rm pole}|^4
=
\frac{\eta^2-1}{\eta^2+1}
=
\tanh\!\left(\frac{1}{\xi_{\rm loc}}\right).
\end{equation}
As \(\eta\to1^+\), \(|q_p|\to1\), \(\xi_{\rm loc}\to\infty\), and
${\rm IPR}_{\rm pole}\to0$, which shows that the boundary
mode and its associated Wigner droplet, delocalizes and merges
with the continuum. 

We emphasize that the droplet should be interpreted as the phase-space
signature of a localized eigenmode of \(H_{\rm eff}\), rather than as a feature of the full physical Wigner function.
For instance, before the wavepacket reaches the boundary, it spreads
essentially ballistically. In that case, although the droplet contribution can be
formally isolated, Eq.~(\ref{WigPole}) also contains continuum and
pole--continuum interference terms, whose coherent sum with the pole
contribution result in a Wigner function exhibiting the characteristic patterns of a free walker. After interaction
with the boundary, absorption and coherent reflection generate
interference fringes and progressive probability depletion. These features
are shown in Fig.~\ref{CV}.

 Since \(W_+\) is constructed from the unnormalized surviving state,
its total weight decreases with the survival probability \(S(t)\).
Thus, it is also useful to define the conditionally normalized
distribution
\begin{equation}
\widetilde W_+(m,k,t)
=
\frac{W_+(m,k,t)}{S(t)},
\end{equation}
which removes the overall probability depletion and isolates the
phase-space structure of the walker conditioned on nonabsorption. As a result, $\widetilde W_+(m,k,t)$ displays the evolving shape and interference
structure of the surviving component more clearly. The normalized counterpart of Fig.~\ref{CV} is provided in\ref{app:normalized_wigner}.
    \section{Conclusions}
    
    We introduced an exactly solvable continuous-time QW with absorption generated by a Lindblad boundary sink. The dynamics exhibits a crossover at \(\kappa/\Omega=1\), separating weak coupling from a strongly dissipative regime in which transport into the boundary becomes dynamically suppressed. This threshold is signaled by the emergence of a boundary-localized non-Hermitian mode, whose Wigner representation forms a droplet exponentially confined near the edge site. Despite the distinct mechanisms governing the weak- and strong-absorption regimes, the total absorption probability becomes invariant under
    \(
    \eta\leftrightarrow\eta^{-1}
    \) as the initial site is moved far from the edge site.
    The Wigner representation further reveals how a boundary sink reshapes the phase-space structure of the walk through interference and probability depletion.  Our framework provides an exactly solvable open-system realization of absorbing QW dynamics in which first-passage statistics arise directly from continuous boundary loss rather than from an externally imposed detection protocol.

   Several directions merit further investigation. The Green-function
approach pursued here extends naturally to other graph topologies:
for a general graph with an absorbing node, the sink again produces a
rank-one non-Hermitian perturbation. The geometry of the graph  enters
through the corresponding unperturbed Green function. Existence
and localization of absorber-induced boundary modes, as well as the
associated first-passage statistics, will depend on the
spectral structure of the underlying graph. By contrast, the simple
threshold and weak--strong duality found here rely on the half-line
geometry, so it is not obvious they persist in more general topologies. Extensions
to multiple absorbers or finite geometries may thus reveal
richer transport and trapping phenomena generated by interference
between boundaries. Another interesting direction is to study the effects of  interactions with an external environment, which is
relevant for realistic transport platforms.  Coupling the walk to bosonic degrees of freedom provides a natural
setting to explore the interplay between coherent propagation,
absorption, and decoherence. More broadly, our absorbing QW model
offers a versatile arena for the study of non-Hermitian physics and
open-system transport beyond measurement-based protocols.
    
  \ack
This work is supported by DOE EXPRESS award No.~DE-SC0024685.
Additional support by DOE award DE-SC0026373 is acknowledged.
The author would like to thank Alejo Nahuel Rossia for a critical reading
of the manuscript, and Manuel Osvaldo C\'aceres and Juan Pablo Rossetti
for fruitful early discussions. The author acknowledges use of ChatGPT-5.6 to assist with mathematical consistency checks, numerical
code development, and editorial tasks.
The author alone is responsible for the claims and content of the manuscript. 

\medskip
\noindent\textbf{Conflict of interest.}
The author declares no competing interests.

\smallskip
\noindent\textbf{Data availability.}
No new data were generated or analysed during this study. The Mathematica
code used to generate the numerical figures and perform the convergence
checks is publicly available at
\url{https://github.com/FranciscoRiberi123/absorbing-quantum-walk}

\smallskip
\noindent\textbf{Ethics statement.}
This work did not involve human participants, human tissue, animals,
or field studies. Ethical approval was therefore not required.

\appendix

\section{Green-function solution of the boundary defect problem}
\label{app:green}

\subsection{Green-function formulation}
To solve the absorbing quantum walk exactly, we compute the Green function of the non-Hermitian tight-binding Hamiltonian governing the surviving walker sector,
\begin{equation}
    H_{\rm eff}
=
H_+
-\frac{i\kappa}{2}\ket{1}\bra{1}.
\end{equation}

We define the Green functions
\(
G_\kappa(z)=(z-H_{\rm eff})^{-1}
\)
and
\(
G_D(z)=(z-H_+)^{-1},
\)
corresponding respectively to the absorbing problem and hard-wall limit with no absorption. We express the absorber as a rank-one perturbation
\(
V_{\rm abs}
=
-\frac{i\kappa}{2}\ket{1}\bra{1}.
\)
The Green functions then satisfy the exact operator relation
\begin{eqnarray}
G_\kappa(z)
&=&
G_D(z)
+
G_D(z)V_{\rm abs}G_\kappa(z).
\label{DysonRankOne}
\end{eqnarray}

\subsection{Exact solution for the rank-one defect}

We consider the matrix elements in the position eigenbasis,
\(
G_\kappa(s,s_0;z)
=
\mel{s}{G_\kappa(z)}{s_0}
\).
Equation ~(\ref{DysonRankOne}) becomes
\begin{eqnarray}
G_\kappa(s,s_0;z)
&=&
G_D(s,s_0;z)
-
\frac{i\kappa}{2}
G_D(s,1;z)\,
G_\kappa(1,s_0;z).
\label{DysonMatrixElement}
\end{eqnarray}

Setting \(s=1\) yields a closed equation for the boundary Green function,
\begin{eqnarray}
G_\kappa(1,s_0;z)
&=&
G_D(1,s_0;z)
-
\frac{i\kappa}{2}
G_D(1,1;z)\,
G_\kappa(1,s_0;z).
\end{eqnarray}
Solving for $G_{\kappa}(1,s_0;z)$, in turn, yields
\begin{eqnarray}
G_\kappa(1,s_0;z)
&=&
\frac{
G_D(1,s_0;z)
}{
1+\frac{i\kappa}{2}G_D(1,1;z)
}.
\label{BoundaryResolvent}
\end{eqnarray}

Substituting Eq.~(\ref{BoundaryResolvent}) into
Eq.~(\ref{DysonMatrixElement}) gives the exact Green function,
\begin{eqnarray}
G_\kappa(s,s_0;z)
&=&
G_D(s,s_0;z)
-
\frac{i\kappa}{2}\,
\frac{
G_D(s,1;z)\,
G_D(1,s_0;z)
}{
1+\frac{i\kappa}{2}G_D(1,1;z)
}.
\label{ExactRankOneGreen}
\end{eqnarray}

\subsection{Full-line lattice Green function}

To compute the half-line propagator, we first determine the Green function of the translationally invariant infinite chain. Diagonalizing the bulk Hamiltonian in momentum space gives the dispersion relation
\(
E(k)=\Omega(1-\cos k),
\)
with Bloch eigenstates \(\ket{k}\propto \sum_{s \in \mathbb Z} e^{iks}\ket{s}\) \cite{nizama2012}. Expanding in this eigenbasis, the corresponding Green function admits the following spectral representation:
\begin{eqnarray}
g_n(z)
&=&
\int_{-\pi}^{\pi}
\frac{dk}{2\pi}
\frac{e^{ikn}}
{z-\Omega(1-\cos k)},
\qquad
n\in\mathbb Z.
\label{gnFourier}
\end{eqnarray}

We introduce
\(
w=e^{ik}
\)
and map the Brillouin zone to the unit circle \(|w|=1\). We may rewrite
Eq.~(\ref{gnFourier}):
\begin{eqnarray}
g_n(z)
&=&
\frac{2}{\Omega}
\oint_{|w|=1}
\frac{dw}{2\pi i}
\frac{w^n}
{
w^2
+
2\frac{z-\Omega}{\Omega}w
+
1
}. \label{gnw}
\end{eqnarray}
It is convenient to define a new variable \(q=q(z)\) implicitly through the relationship
\begin{eqnarray}
z
&=&
\Omega
-
\frac{\Omega}{2}
\left(
q+\frac{1}{q}
\right).
\label{qDefinition}
\end{eqnarray}
The denominator in Eq. (\ref{gnw}) can be factored in terms of $q$:
\begin{eqnarray}
w^2
+
2\frac{z-\Omega}{\Omega}w
+
1
&=&
(w-q)(w-q^{-1}).
\end{eqnarray}
We choose the branch that gives \(q\to0\) as \(z\to\infty\).
Since \(|q|<1\), only the pole at \(w=q\) lies inside the unit circle. The integral can be evaluated by residues:
\begin{eqnarray}
g_n(z)
&=&
\frac{2}{\Omega}
\frac{q^n}{q-q^{-1}},
\qquad n\ge0.
\end{eqnarray}
Using
\(
q-q^{-1}
=
-(1-q^2)/q
\)
and the symmetry \(g_{-n}=g_n\), the 
Green function for the translation invariant quantum walk is:
\begin{eqnarray}
g_n(z)
&=&
-\frac{2}{\Omega}
\frac{q^{|n|+1}}{1-q^2}.
\label{gnClosed}
\end{eqnarray}

\subsection{Half-line Green function and exact boundary solution}

The hard-wall Green function on the half-line follows from the method of images.
Explicitly, substituting Eq.~(\ref{gnClosed}) gives
\begin{eqnarray}
G_D(s,s_0;z)&=&g_{s-s_0}(z)-g_{s+s_0}(z)
\nonumber\\
&=&
-\frac{2}{\Omega(1-q^2)}
\left(
q^{|s-s_0|+1}
-
q^{s+s_0+1}
\right), \qquad
s,s_0\ge1.
\end{eqnarray}
The boundary matrix elements then simplify to:
\begin{equation}
G_D(s,1;z)
=
-\frac{2}{\Omega}q^s,\qquad
G_D(1,s_0;z)
=
-\frac{2}{\Omega}q^{s_0}, \qquad
G_D(1,1;z)
=
-\frac{2}{\Omega}q.    
\end{equation}

Substituting into Eq.~(\ref{ExactRankOneGreen}) yields a compact expression for the denominator:
\begin{eqnarray}
1+\frac{i\kappa}{2}G_D(1,1;z)
&=&
1-i\eta q,
\qquad
\eta=\frac{\kappa}{\Omega},
\end{eqnarray}
and hence the exact Green function for the absorbing half line is:
\begin{eqnarray}
G_\kappa(s,s_0;z)
&=&
-\frac{2}{\Omega(1-q^2)}
\left(
q^{|s-s_0|+1}
-
q^{s+s_0+1}
\right)
-
\frac{2i\eta}{\Omega}
\frac{q^{s+s_0}}
{1-i\eta q}.
\label{ExactResolvent}
\end{eqnarray}

\section{Inverse Laplace transform and Bessel-series propagator}
\label{app:laplace}

\subsection{Inverse transform of the exact resolvent}

Starting from the exact resolvent derived in \ref{app:green}, Eq. (\ref{ExactResolvent}),
the propagator is obtained by inverse Laplace transform,
\begin{eqnarray}
K_\kappa(s,s_0;t)
&=&
\frac{1}{2\pi i}
\int_{\mathcal C}
dz\,
e^{-izt}
G_\kappa(s,s_0;z),
\label{InverseLaplace}
\end{eqnarray}
where \(\mathcal C\) is the inverse-transform contour
running horizontally above the spectrum of \(H_{\rm eff}\):
\begin{eqnarray}
\mathcal C_{\epsilon,R}:\quad
z&=&x+i\epsilon,
\qquad
x:R\to -R,
\end{eqnarray}
with \(R\to\infty\) and \(\epsilon\to0^+\). Equation~(\ref{InverseLaplace}) may be evaluated by standard residue integration after embedding the Bromwich line in a closed contour which may enclose isolated poles corresponding to localized boundary modes. 
The continuum spectrum of the lattice produces a branch cut along the interval
\([0,2\Omega]\), so the contour is taken to wrap around this branch cut through a standard keyhole deformation, and close in the lower half plane. Thus the inverse transform decomposes into branch-cut and pole contributions. Defining $ \lim_{\epsilon \to 0}\Delta G(E)
=
G(E+i \epsilon)-G(E-i\epsilon)
$, we may write:
\begin{eqnarray}
K_\kappa(s,s_0;t)
&=&
-\frac{1}{2\pi i}
\int_0^{2\Omega}
dE\,
e^{-iEt}
\Delta G_\kappa(s,s_0;E)
\nonumber\\
&&+
\sum_{z_p}
{\mathop{\rm Res}_{z=z_p}}
\left[
e^{-izt}G_\kappa(s,s_0;z)
\right],
\label{ContourDecomp}
\end{eqnarray}

It is convenient to break down the result as
\begin{eqnarray}
K_\kappa(s,s_0;t)
&=&
K_D(s,s_0;t)
+
K_{\rm abs}(s,s_0;t),
\end{eqnarray}
where $K_D$ denotes the hard-wall propagator and
\begin{eqnarray}
K_{\rm abs}(s,s_0;t)
&=&
-\frac{2i\eta}{\Omega}
\frac{1}{2\pi i}
\int_{\mathcal C}
dz\,
e^{-izt}
\frac{q^{s+s_0}}
{1-i\eta q}.
\label{KabsDef}
\end{eqnarray}
Fortunately, the inverse transform of $q^m$ may be obtained directly from the known hard-wall propagator, avoiding explicit contour inversion. Using
\begin{eqnarray}
G_D(m,1;z)
&=&
-\frac{2}{\Omega}q^m,
\end{eqnarray}
together with \cite{nizama2012}
\begin{eqnarray}
K_D(m,1;t)
&=&
e^{-i\Omega t}
i^{m-1}
\left[
J_{m-1}(x)
+
J_{m+1}(x)
\right],
\qquad x=\Omega t,
\end{eqnarray}
comparison with Eq.~(\ref{InverseLaplace}) yields the identity
\begin{eqnarray}
\frac{1}{2\pi i}
\int_{\mathcal C}
dz\,
e^{-izt}
q^m
&=&
-\frac{\Omega}{2}
e^{-i\Omega t}
i^{m-1}
\left[
J_{m-1}(x)
+
J_{m+1}(x)
\right].
\label{qInverseIdentity}
\end{eqnarray}
Equation~(\ref{qInverseIdentity}) provides the basic building block for the exact propagator expansion derived below.

\subsection{Weak-absorption regime \texorpdfstring{$|\eta|<1$}{|eta|<1}}

For $|\eta|<1$, the denominator \(1-i\eta q\) appearing in Eq.~(\ref{KabsDef}) admits the convergent expansion
\begin{eqnarray}
\frac{1}{1-i\eta q}
&=&
\sum_{r=0}^{\infty}
(i\eta)^r q^r.
\end{eqnarray}
Therefore
\begin{eqnarray}
\frac{q^N}{1-i\eta q}
&=&
\sum_{r=0}^{\infty}
(i\eta)^r q^{N+r},
\qquad
N=s+s_0.
\end{eqnarray}

Substituting into Eq.~(\ref{ExactResolvent}) and inverting term by term gives the following series representation:
\begin{eqnarray}
K_{\rm abs}(s,s_0;t)
&=&
-\frac{2i\eta}{\Omega}
\sum_{r=0}^{\infty}
(i\eta)^r
\frac{1}{2\pi i}
\int_{\mathcal C}
dz\,e^{-izt}q^{N+r}.
\end{eqnarray}

Using the inversion identity derived above,
\begin{eqnarray}
\frac{1}{2\pi i}
\int_{\mathcal C}
dz\,e^{-izt}q^m
&=&
-\frac{\Omega}{2}
e^{-i\Omega t}
i^{m-1}
\left[
J_{m-1}(x)
+
J_{m+1}(x)
\right],
\end{eqnarray}
with \(m=N+r\), we obtain
\begin{eqnarray}
K_{\rm abs}(s,s_0;t)
&=&
i\eta e^{-i\Omega t}
\sum_{r=0}^{\infty}
(i\eta)^r
i^{N+r-1}
\left[
J_{N+r-1}(x)
+
J_{N+r+1}(x)
\right]
\nonumber\\
&=&
\eta\,i^N e^{-i\Omega t}
\sum_{r=0}^{\infty}
(-\eta)^r
\left[
J_{N+r-1}(x)
+
J_{N+r+1}(x)
\right].
\end{eqnarray}

Combining with the hard-wall contribution yields the exact propagator
\begin{eqnarray}
K_\kappa^{<}(s,s_0;t)
&=&
K_D(s,s_0;t)
\nonumber\\
&&\quad
+
\eta\,i^{s+s_0}e^{-i\Omega t}
\sum_{r=0}^{\infty}
(-\eta)^r
\left[
J_{s+s_0+r-1}(x)
+
J_{s+s_0+r+1}(x)
\right].
\label{WeakPropagator}
\end{eqnarray}

Using the Bessel recurrence relation
\(
J_{m-1}(x)+J_{m+1}(x)=\frac{2m}{x}J_m(x),
\)
this may be written equivalently in the more compact form
\begin{eqnarray}
K_\kappa^{<}(s,s_0;t)
&=&
K_D(s,s_0;t)
+
2\eta\,i^{N}e^{-i\Omega t}
\sum_{r=0}^{\infty}
(-\eta)^r
\frac{N+r}{x}
J_{N+r}(x). \label{Kcompact}
\end{eqnarray}
The series resums repeated returns of the walker to the absorbing boundary.

\subsection{Strong-absorption regime and the boundary pole}

For \(\eta>1\), the geometric expansion used previously breaks down. Moreover, the denominator \(1-i\eta q\) in Eq.~(\ref{KabsDef}) vanishes at
$q_p
=
-\frac{i}{\eta}.$
As a consequence, the closed contour used to evaluate the inverse transform encloses an isolated pole contribution.
The corresponding complex energy follows from
\(
z
=
\Omega
-
\frac{\Omega}{2}
(q+q^{-1})
\).
Substituting \(q=q_p=-i/\eta\), we obtain
\begin{eqnarray}
z_p
&=&
\Omega
-
\frac{i}{2}
\left(
\kappa-\frac{\Omega^2}{\kappa}
\right).
\end{eqnarray}
Evaluating the inverse transform by residues gives the pole contribution
\begin{eqnarray}
K_{\kappa}^{\rm pole}(s,s_0;t)
&=&
(1-q_p^2)\,
q_p^{\,s+s_0-2}\,
e^{-iz_p t}.
\end{eqnarray}
The remaining continuum contribution is obtained after deforming the
inverse-transform contour onto the branch cut. Along the continuum,
\(q=e^{ik}\) and hence \(|q|=1\). Therefore, for \(\eta>1\),
\(|\eta q|>1\), and the denominator admits the convergent inverse-power
expansion
\begin{eqnarray}
\frac{q^N}{1-i\eta q}
&=&
-\sum_{r=1}^{\infty}
i^{-r}\eta^{-r}q^{N-r}.
\end{eqnarray}
Substituting into Eq.~(\ref{KabsDef}) gives
\begin{eqnarray}
K_{\rm abs}^{\rm cont}(s,s_0;t)
&=&
\frac{2i\eta}{\Omega}
\sum_{r=1}^{\infty}
i^{-r}\eta^{-r}
\frac{1}{2\pi i}
\int_{\mathcal C}
dz\,
e^{-izt}
q^{N-r}.
\end{eqnarray}
Here and below, \(\mathcal C\) in the continuum contribution denotes
the contour deformed around the branch cut, with the isolated pole
contribution already removed.
Using the inversion identity
\begin{eqnarray}
\frac{1}{2\pi i}
\int_{\mathcal C}
dz\,
e^{-izt}
q^m
&=&
-\frac{\Omega}{2}
e^{-i\Omega t}
i^{m-1}
\left[
J_{m-1}(x)
+
J_{m+1}(x)
\right],
\qquad x=\Omega t,
\end{eqnarray}
with \(m=N-r\), we obtain
\begin{eqnarray}
K^{\rm cont}_{\rm abs}(s,s_0;t)
&=&
i^{N}e^{-i\Omega t}
\sum_{r=1}^{\infty}
(-1)^{r+1}\eta^{1-r}
\left[
J_{N-r-1}(x)
+
J_{N-r+1}(x)
\right].
\end{eqnarray}
Thus the continuum contribution, $K^{\rm cont}_{\kappa}(s,s_0;t)=K_D(s,s_0;t) +K_{\rm abs}^{\rm cont}(s,s_0;t) $ takes the form
\begin{eqnarray}
K_{\kappa}^{\rm cont}(s,s_0;t)
&=&
K_D(s,s_0;t)
+
2i^{N}e^{-i\Omega t}
\sum_{r=1}^{\infty}
(-1)^{r+1}\eta^{1-r}
\frac{N-r}{x}
J_{N-r}(x),
\qquad
N=s+s_0.
\label{KcontStrong}
\end{eqnarray}
The total propagator immediately follows:
\begin{eqnarray}
K^>_\kappa(s,s_0;t)
&=&
K_{\kappa}^{\rm cont}(s,s_0;t)
+
K_{\kappa}^{\rm pole}(s,s_0;t).
\label{Kkc}
\end{eqnarray}

\section{Equivalence of the weak- and strong-coupling propagator representations}
\label{equivalence}
The weak-coupling representation can be written as
\begin{equation}
K_\kappa^{<}(s,s_0;t)
=
K_D(s,s_0;t)+i^\ell e^{-ix}W(s,s_0;t), \quad
W(s,s_0;t)
=
\eta\sum_{r=0}^{\infty}(-\eta)^r A_{\ell+r}(x),
\end{equation}
where $\ell=s+s_0,$  $ x=\Omega t,$ and we introduced the shorthand $A_m(x)\equiv J_{m-1}(x)+J_{m+1}(x)$.
Similarly, the strong-coupling representation can be separated into a continuum and a pole contribution, 
\begin{eqnarray}
    K_{\kappa}^{>}(s,s_0;t)= K_\kappa^{\rm cont}(s,s_0;t) + K_{\kappa}^{\rm pole}(s,s_0;t), \quad K_{\kappa}^{\rm pole}(s,s_0;t)= (1-q_p^2)\,q_p^{s+s_0-2}e^{-iz_p t}
\end{eqnarray}
where $q_p=-\frac{i}{\eta}$ and
    $z_p
    =
    \Omega
    -
    \frac{i\Omega}{2}
    \left(
    \eta-\eta^{-1}
    \right).
    $
The continuum contribution is, explicitly
\begin{equation}
K_\kappa^{\rm cont}(s,s_0;t)
=
K_D(s,s_0;t)+i^\ell e^{-ix} C(s,s_0;t), \quad C(s,s_0;t)=\sum_{r=1}^{\infty}
(-1)^{r+1}\eta^{1-r}A_{\ell-r}(x),
\end{equation}
Thus, to prove equivalence, it sufices to show 
\begin{eqnarray}
   i^\ell e^{-i x} (W-C)(s,s_0;t)=K_{\kappa}^{\rm pole}(s,s_0;t).
\end{eqnarray}
In $W(s,s_0;t)$, it is convenient to relabel indices, $m=\ell+r$ such that
\begin{equation}
W (s,s_0;t)
=
\sum_{m=\ell}^{\infty}
\eta(-\eta)^{m-\ell}A_m.
\end{equation}
For $C(s,s_0;t)$,  we define instead
$m=\ell-r.$ Using $r=\ell-m$, we obtain
\begin{equation}
C(s,s_0;t)
=
\sum_{m=-\infty}^{\ell-1}
(-1)^{\ell-m+1}\eta^{1-\ell+m}A_m, \qquad \Rightarrow \qquad
-C(s,s_0;t)
= \sum_{m=-\infty}^{\ell-1}
\eta(-\eta)^{m-\ell}A_m
.
\end{equation}
The two sums now fit together as a single summation over all integers:
\begin{equation}
(W-C) (s,s_0;t)
=
\eta(-\eta)^{-\ell}
\sum_{m=-\infty}^{\infty}
(-\eta)^m A_m(x).
\end{equation}
The entire problem reduces to evaluating this bilateral Bessel sum. Using the standard Bessel generating function
\begin{equation}
\sum_{m=-\infty}^{\infty}t^mJ_m(x)
=
\exp\left[
\frac{x}{2}\left(t-\frac1t\right)
\right],
\end{equation}
evaluated at \(t=-\eta\), we obtain
\begin{equation}
\sum_{m=-\infty}^{\infty}(-\eta)^mA_m
=
-\left(\eta+\eta^{-1}\right)
\exp\left[
-\frac{x}{2}(\eta-\eta^{-1})
\right].
\end{equation}

Substitution into our expression for $W-C$ gives
\begin{equation}
(W-C)(s,s_0;t)
=
(-1)^{\ell+1}(1+\eta^2)\eta^{-\ell}
e^{-\frac{x}{2}(\eta-\eta^{-1})}.
\end{equation}
Restoring the common prefactor $i^\ell e^{-ix}$ leads to a closed-form expression for the difference of the weak-coupling representation propagator and the strong-coupling continuum contribution:
\begin{equation}
K_\kappa^{<}(s,s_0;t)-K_\kappa^{\rm cont}(s,s_0;t)
=
(-1)^{\ell+1}i^\ell
(1+\eta^2)\eta^{-\ell}
e^{-ix}
e^{-\frac{x}{2}(\eta-\eta^{-1})}. \label{int}
\end{equation}
It remains to show that Eq.~(\ref{int}) coincides with
\(K_{\kappa}^{\rm pole}\).
The strong-coupling representation is
\begin{equation}
K_{\kappa}^{\rm pole}(s,s_0;t)
=
(1-q_p^2)q_p^{\ell-2}e^{-iz_pt}, \qquad q_p=-\frac{i}{\eta}, \qquad z_p
=
\Omega-\frac{i\Omega}{2}
(\eta-\eta^{-1}).
\end{equation}
Consider its three factors separately.
First,
\begin{equation}
1-q_p^2
=
1-\left(-\frac{i}{\eta}\right)^2
=
1+\frac1{\eta^2}
=
\frac{1+\eta^2}{\eta^2}.
\end{equation}
Second,
\begin{equation}
q_p^{\ell-2}
=
\left(-\frac{i}{\eta}\right)^{\ell-2}
=
(-i)^{\ell-2}\eta^{-(\ell-2)}.
\end{equation}
The phase can be rewritten as
$(-i)^{\ell-2}=(-1)^{\ell+1}i^\ell.$
Finally, expanding $z_p$,
\begin{equation}
e^{-iz_pt}
=
e^{-ix}
e^{-\frac{x}{2}(\eta-\eta^{-1})}.
\end{equation}
Putting everything together,
\begin{equation}
K_{\kappa}^{\rm pole}(s,s_0;t)
=
(-1)^{\ell+1}i^\ell
(1+\eta^2)\eta^{-\ell}
e^{-ix}
e^{-\frac{x}{2}(\eta-\eta^{-1})},
\end{equation}
matching Eq. (\ref{int})  exactly. Equality between $K_{\kappa}^{>}(s,s_0;t)$ and $K^<_{\kappa}(s,s_0;t)$ follows.

\section{Survival probability and absorption asymptotics}
\label{app:survival}

\subsection{Survival probability and first-passage density}

For an initial state localized at site \(s_0\), the survival probability is
\begin{eqnarray}
S(t|s_0)
=
\sum_{s\ge1}
|K_\kappa(s,s_0;t)|^2.
\end{eqnarray}

The corresponding first-passage density is
\begin{eqnarray}
F(t|s_0)
=
-\frac{d}{dt}S(t|s_0)
=
\kappa\,|K_\kappa(1,s_0;t)|^2,
\end{eqnarray}
and thus entirely determined by the boundary propagator.

\subsection{Total absorption probability}
\label{app:pabs}

The exact absorption probability is defined as
\begin{equation}
P_{\rm abs}(s_0;\eta)
=
\kappa\int_0^\infty dt\,
\left|K_\kappa(1,s_0;t)\right|^2 .
\label{eq:pabs_exact_time_finite_s0}
\end{equation}
As the propagator $K_\kappa(1,s_0;t)$ depends on Bessel functions, Eq.~(\ref{eq:pabs_exact_time_finite_s0}) is difficult to evaluate directly. Thus, we use Parseval's identity to rewrite it as an integral in the frequency domain, exploiting the fact that the Fourier transform of the propagator can be expressed in terms of the Green function $G_{\kappa}(1,s_0; z)$, which has a simpler analytic structure.

To make this connection explicit, we first define the causal extension
\begin{equation}
f(t)=\Theta(t)K_\kappa(1,s_0;t).
\end{equation}
Its Fourier transform is
\begin{equation}
\widetilde f(E)
=
\int_{-\infty}^{\infty}dt\,e^{iEt}f(t)
=
\int_0^\infty dt\,e^{iEt}K_\kappa(1,s_0;t).
\end{equation}
This expression can be related directly to the Green function. Introducing a small convergence factor $e^{-\epsilon t}$, with $\epsilon>0$, and using
\begin{equation}
K_\kappa(1,s_0;t)
=
\langle 1|e^{-iH_{\rm eff}t}|s_0\rangle,
\end{equation}
we have
\begin{eqnarray}
\int_0^\infty dt\,e^{iEt-\epsilon t}
K_\kappa(1,s_0;t)
&=&
\left\langle 1\left|
\int_0^\infty dt\,
e^{[i(E-H_{\rm eff})-\epsilon]t}
\right|s_0\right\rangle
\nonumber\\
&=&
i\left\langle 1\left|
(E-H_{\rm eff}+i\epsilon)^{-1}
\right|s_0\right\rangle
\nonumber\\
&=&
i\,G_\kappa(1,s_0;E+i\epsilon).
\end{eqnarray}
The integral converges because the eigenvalues of $H_{\rm eff}$ have non-positive imaginary parts, while $e^{-\epsilon t}$ provides exponential damping. Taking $\epsilon\to0^+$ gives
\begin{equation}
\widetilde f(E)
=
i\,G_\kappa(1,s_0;E+i0^+),
\label{eq:propagator_resolvent_relation}
\end{equation}
and substituting into Parseval's identity yields
\begin{equation}
\int_0^\infty dt\,
|K_\kappa(1,s_0;t)|^2
=
\frac{1}{2\pi}
\int_{-\infty}^{\infty}dE\,
\left|
G_\kappa(1,s_0;E+i0^+)
\right|^2.
\end{equation}
Substituting, in turn, into Eq.~(\ref{eq:pabs_exact_time_finite_s0}) leads to a tractable expression for the absorption probability:
\begin{equation}
P_{\rm abs}(s_0;\eta)
=
\frac{\kappa}{2\pi}
\int_{-\infty}^{\infty}dE\,
\left|
G_\kappa(1,s_0;E+i0^+)
\right|^2.
\label{eq:pabs_parseval}
\end{equation}
The required boundary Green function follows by evaluating Eq.~(\ref{ExactResolvent}) at $s=1$ and $z=E+i0^+$,
\begin{equation}
G_\kappa(1,s_0;E+i0^+)
=
-\frac{2}{\Omega}
\frac{q^{s_0}}{1-i\eta q},
\label{eq:exact_boundary_resolvent_finite_s0}
\end{equation}
with $q\equiv q(E+i0^+)$ determined by the dispersion relation in Eq.~(\ref{qDefinition}). Substituting the exact boundary Green function,
\begin{equation}
P_{\rm abs}
=
\frac{2\eta}{\pi}
\int_{-\infty}^{\infty}
d\varepsilon\,
\frac{|q|^{2s_0}}
{|1-i\eta q|^2},
\qquad
\varepsilon=\frac{E}{\Omega}
=
1-\frac12\left(q+\frac1q\right).
\label{eq:pabs_q_full}
\end{equation}
We now rewrite Eq. (\ref{eq:pabs_q_full}) in a more accessible way. We select the branch of $q(z)$ satisfying $|q(z)|<1$ for $\Im z>0$, as required for the Green function to decay at large distance. The real-$E$ axis naturally separates into the interval \(0<E<2\Omega\) and the two exterior intervals. 

Inside \(0<E<2\Omega\), the boundary value $q(E+i 0^+)$ of this branch satisfies $|q|=1$. We parametrize $q=e^{ik}$, so that
\begin{equation}
E=\Omega(1-\cos k),
\qquad
0<k<\pi,
\qquad
dE=\Omega\sin k\,dk,
\end{equation}
and
\begin{equation}
|1-i\eta e^{ik}|^2
=
1+\eta^2+2\eta\sin k.
\end{equation}
Since $|q|^{2s_0}=1$, this contribution can be expressed as an $s_0$-independent integral:
\begin{equation}
P_{\rm in}(\eta)
=
\frac{2\eta}{\pi}
\int_0^\pi
dk\,
\frac{\sin k}
{1+\eta^2+2\eta\sin k}.
\label{eq:pband}
\end{equation}

For energies outside $[0,2\Omega]$, the solutions no longer satisfy $|q|=1$. Instead, for \(E<0\), \(q\in(0,1)\), while for \(E>2\Omega\), \(q\in(-1,0)\). It is convenient to express these terms using $q$ as the integration variable, rather than $E$. We have
\begin{equation}
|1-i\eta q|^2=1+\eta^2q^2, \qquad
E=
\Omega \left[1-\frac12\left(q+\frac1q\right) \right], \qquad
\left|\frac{dE}{dq}\right|
=
\frac{\Omega}{2}
\frac{1-q^2}{q^2}.
\end{equation}
Changing variables $q\mapsto-q$, it is easy to see that both outer contributions are equal, as the integrand depends only on \(q^2\). We write their sum as
\begin{equation}
P_{\rm out}(s_0;\eta)
=
\frac{2\eta}{\pi}
\int_0^1 dq\,
\frac{
q^{2s_0-2}(1-q^2)
}{
1+\eta^2q^2
}.
\label{eq:pout}
\end{equation}
Combining Eqs.~(\ref{eq:pband}) and (\ref{eq:pout}), we obtain the exact finite-$s_0$ absorption probability:
\begin{eqnarray}
P_{\rm abs}(s_0;\eta)
&=&
\frac{2\eta}{\pi}
\int_0^\pi dk\,
\frac{\sin k}
{1+\eta^2+2\eta\sin k}
\nonumber\\
&&+
\frac{2\eta}{\pi}
\int_0^1 dq\,
\frac{
q^{2s_0-2}(1-q^2)
}{
1+\eta^2q^2
}.
\label{eq:pabs_exact_finite_s0}
\end{eqnarray}

The first term is independent of the initial position $s_0$ and invariant under the $\eta\to \eta^{-1}$ transformation. The second term gives a finite-distance correction that is not invariant under $\eta \to \eta^{-1}$. As $s_0$ increases, the integrand becomes increasingly concentrated near $q=1$, since
\begin{equation}
q^{2s_0-2}\to0,
\qquad
0\le q<1.
\end{equation}
To determine how rapidly this contribution vanishes, we change variables $q=1-\frac{x}{2s_0}$ and rewrite Eq.~(\ref{eq:pout}) as
\begin{eqnarray}
P_{\rm out}
&=&
\frac{2\eta}{\pi}\frac{1}{2s_0}
\int_0^{2s_0}dx\,
\frac{
\left(1-\frac{x}{2s_0}\right)^{2s_0-2}
\left[1-\left(1-\frac{x}{2s_0}\right)^2\right]
}{
1+\eta^2\left(1-\frac{x}{2s_0}\right)^2
}
\nonumber\\
&\simeq&
\frac{\eta}{\pi(1+\eta^2)s_0^2}
\int_0^\infty dx\,x e^{-x}
\left[
1+
\frac{1}{s_0}
\left(
\frac{3x}{4}
-\frac{x^2}{4}
+
\frac{\eta^2x}{1+\eta^2}
\right)
+
O(s_0^{-2})
\right]
\nonumber\\
&=&
\frac{\eta}{\pi(1+\eta^2)s_0^2}
\left[
1+
\frac{2\eta^2}{(1+\eta^2)s_0}
+
O(s_0^{-2})
\right].
\end{eqnarray}
In the second line, we have extended the integration limit to infinity, as the limiting integrand is exponentially suppressed at large $x$, and expanded the integrand in powers of $1/s_0$. Thus, the finite-$s_0$ correction vanishes as $s_0^{-2}$. Moreover, its leading coefficient is itself invariant under $\eta\longrightarrow\eta^{-1}$, so deviations from weak--strong duality appear only at order $s_0^{-3}$. Thus, as we move the initial condition away from the absorber, the corrections to this symmetry vanish. As shown in Fig.~\ref{SF} in the main text, already for $s_0=8$ the difference between reciprocal absorption probabilities is essentially negligible.

\section{Exact Wigner representation}
\label{app:wigner}

\subsection{Definition}

To characterize the phase-space structure of the absorbing QW, we introduce the doubled-lattice Wigner function \cite{bizarro1994,hinarejos2012},
\begin{eqnarray}
W_+(m,k,t)
=
\frac{1}{2\pi}
\sum_{n=1}^{m-1}
\rho^{(+)}_{n,m-n}(t)\,
e^{-i(2n-m)k},
\label{WignerDefinitionApp}
\end{eqnarray}
where \(m\ge2\), \(k\in[-\pi,\pi]\), and
\(\rho^{(+)}(t)\) is the unnormalized density matrix in the surviving
walker sector. The restricted summation range follows from the half-line geometry:
\(
\rho^{(+)}_{n,m-n}(t)\neq0
\)
only if \(n\ge1\) and \(m-n\ge1\).

For an initially localized walker at \(s_0\), the reduced state matrix elements are
\(
\rho^{(+)}_{ss'}(t)
=
K_\kappa(s,s_0;t)
K_\kappa^*(s',s_0;t),
\)
so the Wigner distribution adopts a simple form:
\begin{eqnarray}
W_+(m,k,t)
=
\frac{1}{2\pi}
\sum_{n=1}^{m-1}
K_\kappa(n,s_0;t)
K_\kappa^*(m-n,s_0;t)
e^{-i(2n-m)k}.
\label{ExactWignerRepresentation}
\end{eqnarray}
Integrating over \(k\) and summing over the doubled-lattice coordinate
recovers the total weight of the surviving sector:
\begin{eqnarray}
\sum_{m=2}^{\infty}
\int_{-\pi}^{\pi} dk\,
W_+(m,k,t)
&=&
\sum_{s\ge1}\rho^{(+)}_{ss}(t)=
S(t).
\label{eq:wigner_survival_weight}
\end{eqnarray}
Thus \(W_+\) retains both the phase-space structure of the surviving
walker and its decreasing survival probability.

\subsection{Weak dissipation regime}

When \(|\eta|<1\), the propagator splits naturally into hard-wall and absorber contributions, Eq.~(\ref{Kcompact}). Removing the common phase, we write
\begin{eqnarray}
K_\kappa(s,s_0;t)
=
e^{-i\Omega t}
\left[
\Phi_s^D(t)
+
\Phi_s^B(t)
\right],
\end{eqnarray}
where
\begin{eqnarray}
\Phi_s^D(t)
&=&
i^{s-s_0}J_{s-s_0}(x)
-
i^{s+s_0}J_{s+s_0}(x),
\nonumber\\
\Phi_s^B(t)
&=&
\eta\, i^{s+s_0}
\sum_{r=0}^{\infty}
(-\eta)^r
\left[
J_{s+s_0+r-1}(x)
+
J_{s+s_0+r+1}(x)
\right],
\end{eqnarray}
with \(x=\Omega t\). The common phase \(e^{-i\Omega t}\) cancels in the bilinear product entering the Wigner function. Substituting into Eq.~(\ref{ExactWignerRepresentation}), we obtain:
\begin{eqnarray}
W_+(m,k,t)
=
\frac{1}{2\pi}
\sum_{n=1}^{m-1}
\left[
\Phi_n^D+\Phi_n^B
\right]
\left[
\Phi_{m-n}^{D*}+\Phi_{m-n}^{B*}
\right]
e^{-i(2n-m)k}.
\label{Wigdecomp}
\end{eqnarray}
Expanding the product yields the decomposition given in the main text:
\begin{eqnarray}
W_+
=
W_{DD}
+
W_{DB}
+
W_{BD}
+
W_{BB}.
\label{WignerDecomposition}
\end{eqnarray}

\subsection{Strong dissipation regime}

For \(\eta>1\), the propagator separates into continuum and pole pieces,
Eq.~(\ref{Kkc}),
\begin{eqnarray}
K^>_\kappa(s,s_0;t)
=
K_{\kappa}^{\rm cont}(s,s_0;t)
+
K_{\kappa}^{\rm pole}(s,s_0;t),
\end{eqnarray}
with
\begin{eqnarray}
K_{\kappa}^{\rm pole}(s,s_0;t)
=
(1-q_p^2)q_p^{s+s_0-2}e^{-iz_p t},
\qquad
q_p=-\frac{i}{\eta}.
\end{eqnarray}
Substitution into the half-line Wigner definition gives the corresponding
four-term decomposition:
\begin{eqnarray}
W_+
=
W_{\rm cont,cont}
+
W_{\rm cont,pole}
+
W_{\rm pole,cont}
+
W_{\rm pole,pole},
\end{eqnarray}
where
\begin{eqnarray}
W_{\rm cont,cont}(m,k,t)
&=&
\frac{1}{2\pi}
\sum_{n=1}^{m-1}
K_{\kappa}^{\rm cont}(n,s_0;t)
K_{\kappa}^{\rm cont\,*}(m-n,s_0;t)
e^{-i(2n-m)k},
\nonumber\\
W_{\rm cont,pole}(m,k,t)
&=&
\frac{1}{2\pi}
\sum_{n=1}^{m-1}
K_{\kappa}^{\rm cont}(n,s_0;t)
K_{\kappa}^{\rm pole\,*}(m-n,s_0;t)
e^{-i(2n-m)k},
\nonumber\\
W_{\rm pole,cont}(m,k,t)
&=&
W_{\rm cont,pole}^*(m,k,t).
\end{eqnarray}
The pure pole contribution can be evaluated explicitly as the corresponding wavefunction is geometric. Defining
\begin{eqnarray}
\Gamma_p
=
\kappa-\frac{\Omega^2}{\kappa},
\qquad
{\rm Im}\,z_p=-\frac{\Gamma_p}{2},
\end{eqnarray}
we have
\begin{eqnarray}
W_{\rm pole,pole}(m,k,t)
&=&
\frac{1}{2\pi}
\sum_{n=1}^{m-1}
K_{\kappa}^{\rm pole}(n,s_0;t)
K_{\kappa}^{\rm pole\,*}(m-n,s_0;t)
e^{-i(2n-m)k}.
\end{eqnarray}
Substituting the pole wavefunction gives
\begin{eqnarray}
W_{\rm pole,pole}(m,k,t)
=
\frac{|1-q_p^2|^2}{2\pi}
e^{-\Gamma_p t}
q_p^{s_0-2}
(q_p^*)^{m+s_0-2}
e^{imk}
\frac{\alpha-\alpha^m}{1-\alpha},
\label{WppClosed}
\end{eqnarray}
with $\alpha=\frac{q_p}{q_p^*}e^{-2ik}$.
Replacing the explicit value \(q_p=-i/\eta\), we have
\(
\alpha=-e^{-2ik},
\)
and the finite geometric series may be written equivalently as
\begin{eqnarray}
W_{\rm pole,pole}(m,k,t)
&=&
\frac{(1+\eta^{-2})^2}{2\pi}
e^{-\Gamma_p t}
\eta^{-(m+2s_0-4)}
\sum_{n=1}^{m-1}
\cos\!\left[
\frac{\pi m}{2}
+
\pi n
+
(m-2n)k
\right].
\label{WppReal}
\end{eqnarray}
The magnitude of the pole envelope scales as
\begin{eqnarray}
\left|
q_p^{s_0-2}
(q_p^*)^{m+s_0-2}
\right|
=
|q_p|^{m+2s_0-4}
=
\eta^{-(m+2s_0-4)}.
\end{eqnarray}
The factor \(\eta^{-m}\) controls the spatial confinement of the
pole-only Wigner contribution, while
\(\eta^{-2(s_0-2)}\) reflects the exponentially decreasing overlap of an
initial state localized farther from the boundary with the pole mode.
As a consequence, this phase-space contribution is exponentially confined near the absorbing boundary. As the pole amplitude obeys
\(
|K_{\rm pole}(s,s_0;t)|
\propto
e^{-(s-1)/\xi_{\rm loc}},
\)
we define the pole-mode amplitude localization length as
\begin{eqnarray}
\xi_{\rm loc}
=
\frac{1}{\ln\eta}.
\end{eqnarray}
Accordingly, the pole probability decays as
\(e^{-2(s-1)/\xi_{\rm loc}}\).

\begin{figure*}[t!]
    \centering
    \includegraphics[width=\textwidth]{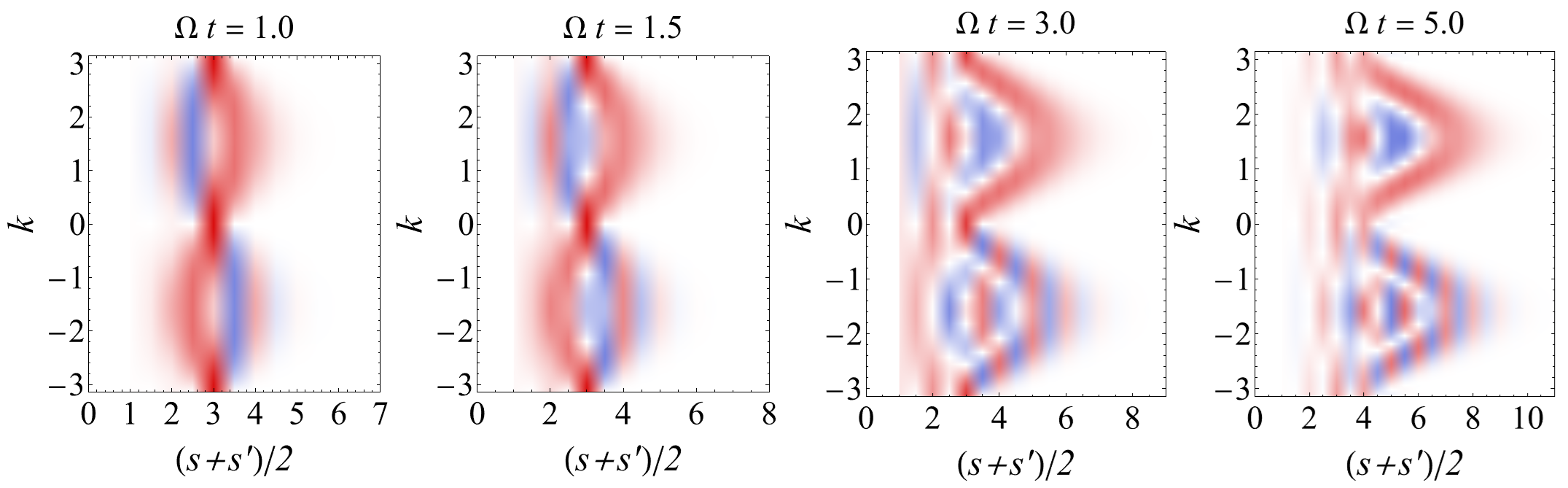}
    \caption{
    \textbf{Conditionally normalized phase-space dynamics.}
    Same parameters and times as in Fig.~\ref{CV}, but showing
    $\widetilde W_+(m,k,t)=W_+(m,k,t)/S(t)$.
    Normalization by the survival probability removes the overall
    depletion caused by absorption and isolates the phase-space
    structure of the walker conditioned on nonabsorption.
    The surviving component continues to spread into the bulk and
    develops boundary-induced interference fringes.
    }
    \label{fig:normalized_wigner}
\end{figure*}

 \subsection{Conditionally normalized Wigner function}
\label{app:normalized_wigner}

The Wigner function shown in Fig.~\ref{CV} is constructed from the
unnormalized surviving state. Thus, it contains information about
both its phase-space structure and its decreasing survival probability.
To separate these effects, we define the normalized conditional state
\begin{equation}
\widetilde\rho^{(+)}(t)
=
\frac{\rho^{(+)}(t)}{S(t)}.
\end{equation}
By linearity of the Wigner transform, its corresponding phase-space
distribution is
\begin{equation}
\widetilde W_+(m,k,t)
=
\frac{W_+(m,k,t)}{S(t)}.
\label{eq:conditional_wigner}
\end{equation}
Using Eq.~(\ref{eq:wigner_survival_weight}), the normalized Wigner
function satisfies
\begin{equation}
\sum_{m=2}^{\infty}
\int_{-\pi}^{\pi}dk\,
\widetilde W_+(m,k,t)
=
1.
\end{equation}
Equivalently, $\widetilde W_+$ is constructed from
the normalized conditional state
\begin{equation}
|\widetilde\psi(t)\rangle
=
\frac{|\psi(t)\rangle}{\sqrt{S(t)}}.
\end{equation}
Equation (\ref{eq:conditional_wigner}) describes the phase-space distribution of the walker
conditioned on nonabsorption.

Figure~\ref{fig:normalized_wigner} shows the conditionally normalized
counterpart of Fig.~\ref{CV}. Removing the overall survival weight
isolates the evolving phase-space structure of the surviving component.
This makes clear that the decreasing magnitude of the unnormalized
Wigner function in Fig.~\ref{CV} reflects probability loss rather than
the disappearance of the underlying phase-space structure. The surviving
walker continues to spread into the bulk and develops pronounced
interference fringes due to scattering from the boundary.

\vspace{9mm}

\providecommand{\newblock}{}

    \end{document}